\documentclass[fleqn,usenatbib]{mnras}

\usepackage{newtxtext,newtxmath}
\usepackage[T1]{fontenc}
\usepackage{ae,aecompl}

\usepackage{amsmath}
\usepackage{amssymb}
\usepackage{graphicx}
\usepackage{epstopdf}

\title[Dimensionality of Trappist-1]{Dimensionality and integrals of motion of the Trappist-1 planetary system}

\author[J. Flo\ss{} et al.]{
Johannes Flo\ss{},$^{1}$
Hanno Rein,$^{2,3}$
and Paul Brumer$^{1}$
\\
$^{1}$Chemical Physics Theory Group, Department of Chemistry, and Centre for Quantum Information and Quantum Control, University of Toronto, Toronto,\\ ON M5S 3H6, Canada\\
$^{2}$Department of Physical \& Environmental Sciences, University of Toronto at Scarborough, Toronto, ON M1C 1A4, Canada\\
$^{3}$Department of Astronomy \& Astrophysics, University of Toronto, Toronto, ON M5S 3H4, Canada}

\date{Accepted XXX. Received YYY; in original form ZZZ}

\pubyear{2018}

\begin{document}
\label{firstpage}
\pagerange{\pageref{firstpage}--\pageref{lastpage}}
\maketitle

\begin{abstract}
The number of isolating integrals of motion of the Trappist-1 system -- a late M-dwarf orbited by seven Earth-sized planets -- was determined numerically, using an adapted version of the correlation dimension method.
It was found that over the investigated time-scales of up to 20~000 years the number of isolating integrals of motion is the same as one would find for a system of seven non-interacting planets -- despite the fact that the planets in the Trappist-1 system are strongly interacting.
Considering perturbed versions of the Trappist-1 system shows that the system may occupy an atypical part of phase-space with high stability.
These findings are consistent with earlier studies.
\end{abstract}


\section{Introduction}

The planetary system of Trappist-1 consists of at least seven planets orbiting a dwarf star.
The planets have roughly equal mass and orbital periods between one and nineteen days.
A remarkable feature of their orbits is that the periods are all in near-resonance~\citep{gillon17,luger17}.
The current dynamical configuration of the system may help us understand its formation history \citep{TamayoReinPetrovichMurray2017}.
This article sheds more light on the dynamics of the Trappist-1 planetary system by focusing on determining the number of isolating integrals of motion.
(We follow the definition by~\citet{binney08} for the term integral of motion. However, for ease of reading we drop the term ``isolating'' in the remainder of the article.)

Knowledge of the number of conserved quantities in a physical system of interest can greatly help to understand the system dynamics:
It can reveal whether different degrees of freedom are coupled or not, or whether the dynamics are regular or chaotic.
And, significantly, it can tell, when detailed integrals of motion are sought,
whether all have been found. 
A prominent example for the latter problem is the search for a third integral of motion in celestial dynamics~\citep{contopoulos63}.
Another, simple example, is a single planet orbiting a central star.
If it is indeed a Kepler problem, i.e. just two bodies coupled by an $1/r$ potential, the system would have five integrals of motion: energy, three components of the angular momentum, and one additional integral arising from the Laplace-Runge-Lenz-vector.
If now in an experimentally observed orbit of the single planet only four integrals of motion are found, one can directly deduce that a disturbing interaction, e.g. caused by a second planet or solar winds, is present.

Finding the number of integrals of motion is in general not a trivial task.
Using analytical methods it usually equates to finding all integrals of motion themselves;
practically impossible for all but the simplest systems.
Consequently, a number of numerical methods have been proposed~\citep{stine83,carnevali84,martens84,barnes01,carpintero08}.
They all rely on the fact that a trajectory in a classical conservative system traces out a surface $S$ of dimension $\nu$.
This dimension is related to the number $C$ of the conserved quantities via
\begin{equation}
\nu=2d-C \,,
\label{eq.dim}
\end{equation}
where $d$ is the number of degrees of freedom and $2d$ is the dimension of the embedding phase-space.
The idea behind the proposed schemes is to numerically determine $\nu$ and thereby $C$.
The approach applied here is based on the correlation dimension methodology
introduced by~\citet{grassberger83a,grassberger83b}.
Since this method was developed for characterizing strange attractors in non-conservative systems,
a number  of corrections were introduced to adapt it for the conservative system considered 
in this study.
A detailed description of the modified method is presented in Sec.~\ref{sec.methods}.

The main part of this article is Sec.~\ref{sec.results}, which describes
 possible scenarios one could encounter in a planetary system, and provides results
 for the dimensional analysis of both the Trappist-1 planet as
 well as derivatives of this system.
Finally, a tentative interpretation of the obtained data is provided.

\section{Methods}
\label{sec.methods}

To obtain the dimension $\nu$ of the phase-space structures of the planetary system of Trappist-1, we
utilise a modified version of the correlation dimension method
\citep{grassberger83a,grassberger83b}.
A brief review of the method is given below.
For a more detailed description, the reader may refer to the original works.
A good introduction to its numerical implementation is provided by \citet{sprott03}.

Consider a system with $d$ degrees of freedom and thus a $2d$-dimensional phase-space formed by the coordinates $q_i$ and momenta $p_i$.
A phase-space trajectory $\mathbfit{X}(t)=(q_1(t),q_2(t),...,q_d(t),p_1(t),p_2(t),...,p_d(t))$ traces out a $\nu$-dimensional surface $S$ in phase-space.
To numerically determine $\nu$  one has to obtain a set $\{\mathbfit{X}_n\}$ of points lying on $S$.
A simple way is to propagate the trajectory and take a point at regular intervals $\tau$, i.e. $\mathbfit{X}_n=\mathbfit{X}(n\tau)$.
One then calculates the correlation sum $C(l)$ which counts the number of all pairs $(\mathbfit{X}_m,\mathbfit{X}_n)$ with a distance smaller than $l$:
\begin{equation}
C(l)=A\sum_{\substack{m,n=1\\|m-n|>s}}^{N} \Theta\left(l-|\mathbfit{X}_m-\mathbfit{X}_n| \right) \,.
\label{eq.correlationsum}
\end{equation}
Here, $N$ is the number of phase-space points in the set $\{\mathbfit{X}_n\}$, $\Theta$ is the Heaviside step function, and $A$ is a normalisation factor.
The parameter $s$ provides a minimal temporal separation between points of the trajectory:
To avoid short-time temporal correlations influencing the algorithm~\citep{grassberger91}, only phase-space point pairs $(\mathbfit{X}_m,\mathbfit{X}_n)$ 
with a temporal separation $\Delta t = |m-n| \tau$ larger than $s\tau$ are included in the correlation sum. For the planetary system of Trappist-1 a value of $s \tau=1$ was found to be sufficient to avoid temporal correlations; to include a safety margin $s\tau=10$ was used in this study. Note that here and in the remainder of this work, all times are given in $\mathrm{yr}/2\pi \approx 58~\mathrm{d}$, if not mentioned otherwise.

Before calculating $C(l)$, coordinates and momenta are scaled to obtain a metric of the phase-space.
In particular, each coordinate and momentum is scaled independently such that its extension is equal to unity:
\begin{equation}
\tilde x_i(t)= x_i(t)/\left[\max(x_i(t)) - \min(x_i(t)) \right] \,.
\end{equation}
Here, the tilde denotes the scaled quantity, and $x$ stands for either $q$ or $p$.
Scaling was not done when $\max(x_i(t))-\min(x_i(t))$ was several orders of magnitude smaller than for the other quantities, as in this case it is assumed that $x_i$ is actually conserved, with any differences being due to numerical errors.

The original algorithm by~\citet{grassberger83a,grassberger83b} puts a higher emphasis on phase-space regions that are visited more frequently by the trajectory, and thus the algorithm may yield a lower dimension for the surface $S$ than its geometric dimension. However, in order to obtain the number of integrals of motion, the geometric dimension of $S$ is needed. Thus, in this work the algorithm is modified and an additional scaling following the idea of Monte Carlo sampling on an energy surface \citep{bunker62} is introduced: individual points were removed from the set $\{\mathbfit{X}_n\}$ to ensure a uniform coverage of the surface $S$ by the set.
A point $\mathbfit{X}_n$ on the trajectory was only included in the set if the phase-space velocity $d\mathbfit{X}/dt$ at this point was larger than a random number $r$ equally distributed between $0$ and the largest measured phase-space velocity.

The relation between $C(l)$ and the dimension $\nu$ is given, for sufficiently small $l>0$, by
\begin{equation}
C(l)\propto l^{\nu} \,.
\label{eq.Cproportionality}
\end{equation}
Using Eq.~\eqref{eq.Cproportionality}, the dimension $\nu$ can be obtained numerically by a linear fit of $\log C$ over $\log l$.
For this purpose, a least squares linear fit for $\log (C)$ versus $\log (l)$ is done on an interval $\log (l_0) < \log (l) < \log (l_0) + 0.5$, yielding $\nu(l_0)$ and the square error $\Delta(l_0)$ within this interval.
The parameter $l_0$ is then scanned within an interval $[l_1,l_2]$ of interest, and the value $l_0$ for which $\Delta(l_0)$ is minimal is chosen to provide the final value for $\nu$.

The initial conditions $\mathbfit{X}(t=0)$ for the Trappist-1 system are one of the successful cases from~\cite{TamayoReinPetrovichMurray2017,tamayo_daniel_2017_496153}; the particular data set used is IC0K1.2521e+02mag1.3982e-01.
They thus represent one realisation of the system that can be explained by a planet migration scenario.
The system was then propagated using the \texttt{IAS15} integrator of the \texttt{REBOUND} code~\citep{rebound12,rebound15} to obtain the phase-space points $\mathbfit{X}_n=\mathbfit{X}(t_0+n\tau)$.
Here, the parameter $t_0=0$ defines the time the set $\{\mathbfit{X}_n\}$ is taken from.
Changing $t_0$ allows us to observe a possible change of the dimension over time.

\section{Dimensional analysis of the Trappist-1 planetary system}
\label{sec.results}

\subsection{General remarks}

Before doing a dimensional analysis of the \mbox{Trappist-1} planetary system,
 it is useful to consider possible scenarios that might be confronted.
The system, which consists of eight bodies (seven planets and one central star),
forms a 48-dimensional phase-space.
It can easily be seen that there are at least ten integrals of motion:
 conservation of linear momentum associated with the centre of mass (six constants), 
 the angular momentum (three constants) and the total energy (one constant).
However, there may well be be up to 37 additional integrals of motion. That is, the dimension $\nu$ can lie between one and 38.

One extreme would be a dimension of $\nu=38$, i.e. no additional integrals of motion.
This would most likely imply ergodic and thus chaotic dynamics on the 38-dimensional manifold, rendering this scenario very unlikely on short timescales:
If the system was ergodic, the trajectory would eventually visit each part of the 38-dimensional surface, including regions where one or more planets reach escape velocity -- the Trappist\nobreakdash-1 planetary system would have broken apart long ago and never been observed.
There is however one exception where such a high dimension together with stability is possible:
If the system lies in a region of phase-space that is closely confined by a barrier, not allowing the trajectory to explore the whole 38-dimenisonal manifold in a reasonable time span.

At the opposite extreme is a scenario in which the seven planets are so weakly interacting with one another that this interaction can be neglected.
In this case each planet would orbit the central star on a Keplerian orbit, and thus the whole system would display seven-dimensional dynamics (one free dimension for the orbit of each planet).
This case also seems rather unlikely considering how close the planets orbits' lie to one another~\citep[see e.g.][]{TamayoTriaudMenouRein2015}.

It is even possible to find a lower dimension than that of the independent planet scenario:
If the planets' orbits become coupled, additional integrals of motion could arise, leading to dynamics of even less than seven dimensions.
Considering that the near resonance found for the orbits of the Trappist-1 planetary system are unlikely to be sheer coincidence, this scenario does not seem completely unlikely.

Note that our application of the 
Procaccia-Grassberger methodology provides insight into the integrals of motion that are confined to the phase
space region covered by the trajectory from which points are sampled, and on the scale sampled 
by $l$. Hence,  the dimension of the phase-space dynamics of a system can look different on different length scales:
When calculating the dimension $\nu$ using $C(l)$, one might find a dimension $\nu_{\text{A}}$ for $l< 10^{-4}$ and a larger dimension $\nu_{\text{B}}$ for $l\approx10^{-2}$.
This implies that the surface $S$ has the dimension $\nu_{\text{A}}$ (since the actual dimension is found at the limit of small $l$), but it appears to have the dimension $\nu_{\text{B}}$ when looked at using length scales of $l>10^{-2}$.
As a model example, consider a dust particle in a river:
It surely follows a six-dimensional trajectory in phase-space, undergoing Brownian motion in the water.
Yet, when looking at the dynamics using the entire river as the size scale, one will only see the dust
particle moving on a one-dimensional trajectory towards the sea, all the Brownian motion is lost due 
to the low resolution coarse-graining. Which dimension is important depends on the question of interest:
If one wants to investigate the Brownian motion, then surely the 
microscopic, six-dimensional dynamics is appropriate. However,
if one is interested in the gross flow of dust from the river to the sea, then 
the macroscopic one-dimensional dynamics is appropriate.

\subsection{The Trappist-1 planetary system}

\begin{figure}
\includegraphics[width=\linewidth]{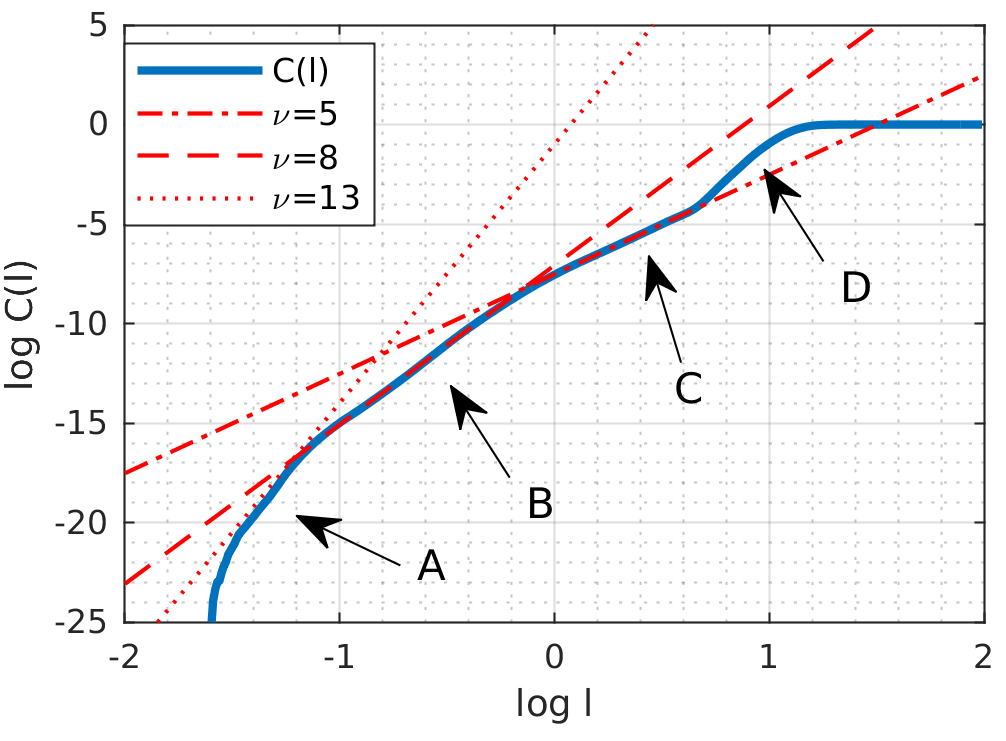}
\caption{
\label{fig.500}
Double logarithmic plot of $C(l)$ over $l$ for the Trappist-1 planetary system, with parameters $t_0=0$, $N=500~000$, $\tau=0.1$, and $s=100$ (see text for a definition of the parameters).
The present situation is found for most sets of parameters $t_0$, $N$, and $\tau$:
Four more or less linear regions A, B, C, and D.
The dotted and dashed lines are curves linear in $\log l$ with integer slope $\nu$, and are there to guide the eye.
}
\end{figure}

In Figure~\ref{fig.500} the correlation sum $C(l)$ is shown for the Trappist-1 planetary system, calculated with typical numerical parameters.
The start time for the sample set is $t_0=0$. 
 The size of the set of phase-space points $\{\mathbfit{X}_n\}$ is $N=500~000$,
a size found to be sufficient to lead to convergence of $C(l)$ for at least $l > \exp(-1)$.
The temporal distance between successive points $\mathbfit{X}_n$ and $\mathbfit{X}_{n+1}$ is chosen as $\tau\approx0.1$~\footnote{The time-step before removing points to achieve a uniform coverage of the surface is set to $\tau=0.1$; the actual average time-step is about 5\% larger.}, thus being on the order of the orbiting periods of the planets.
The total time $T_{\mathrm{tot}}$ covered by the set $\{\mathbfit{X}_n\}$ is then
$T_{\mathrm{tot}}=N\tau=50~000$, approximately 8000 years or 150~000 orbits for the outermost planet.

The graph displayed in Fig.~\ref{fig.500} reveals four regions of essentially linear behaviour in $\log C(l)$; for further use these regions are labelled  A, B, C, and D.
Also shown are straight lines of integer slope that approximately fit $C(l)$ in the regions A, B, and C.

The positions of the regions B,C, and D shift only slightly ($\pm 0.2$) along the $\log l$ axis for different parameters $\tau$, $N$, and $t_0$.
In region A, on the other hand, the curve is not linear for all values of the numerical parameters, and furthermore even for the linear cases a large variety of values for the slope $\nu$ is found.
As it is closest to the limit of $l\to0$, region A might provide the actual dimension of the system, 
but the large noise in this region means that $N$, and possibly the precision of the integrator as well,
is not sufficient to obtain reliable data from this region.
Region D yields the same value $\nu_{\text{D}}=10\pm1$ for most parameters.
Yet, it is at best linear over a very short interval and is also close to the saturation [$C(l)=1$], and thus may be an artifact.
In the following therefore only regions B and C are considered.

\begin{figure}
\includegraphics[width=\linewidth]{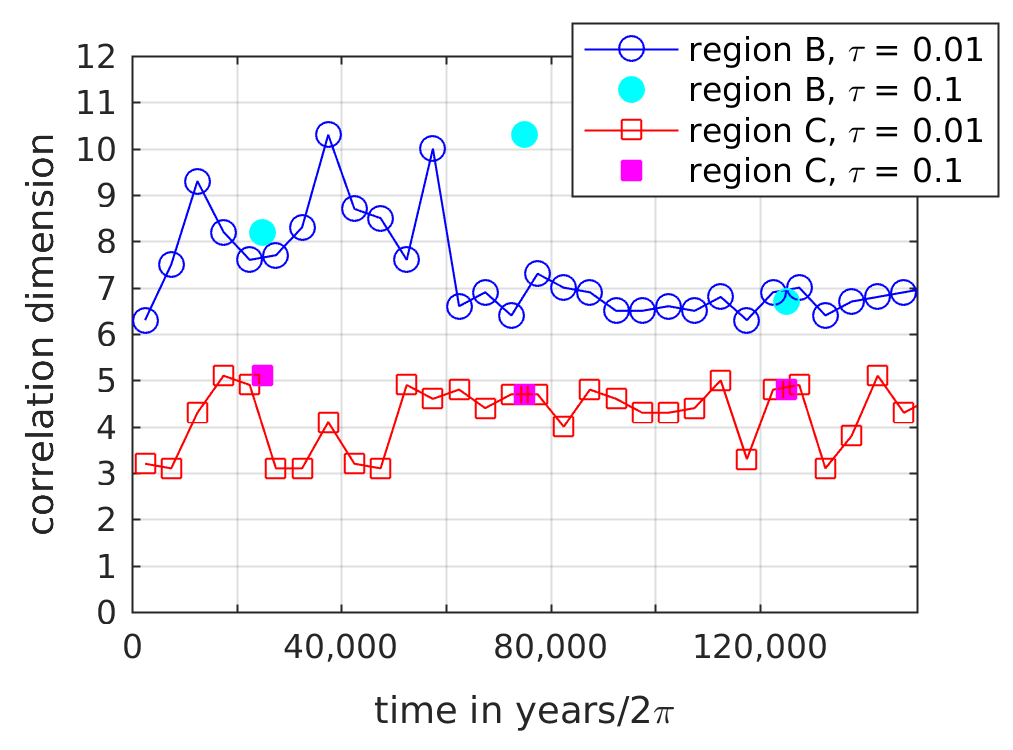}
\caption{
\label{fig.to}
Phase-space dimension of the trajectory of the planetary system of Trappist-1 in the regions B and C at different times.
All markers display the result for sample sizes of $N=500~000$ points on the trajectory.
For the hollow markers, the sample is taken over a time of $T_{\mathrm{tot}}=5000$,  for the full markers $T_{\mathrm{tot}}=50~000$. 
The horizontal axis is the time at the centre of the sampling.
}
\end{figure}

 The dimension of the regions B and C is shown in Fig.~\ref{fig.to} as a function of the time $t_0$ at which the sample set $\{\mathbfit{X}_n\}$ is taken.
In particular, for each value of $t_0$, the trajectory is propagated from the initial state
at $t=0$ until $t_0$, and from there on the phase-space points are sampled; 
i.e., $\mathbfit{X}_n=\mathbfit{X}(t_0+n\tau)$.
The sampling is done with two different values of $\tau$:
The solid lines show the cases for points sampled at a rate of $\tau=0.01$ (500\~000 points are sampled, leading to a set $\{\mathbf{X}_n\}$ obtained over $T_{\mathrm{tot}}\approx 5000 \approx 800~\mathrm{yr}$), the single (filled) markers show longer sampling times with $\tau=0.1$ ($T_{\mathrm{tot}} \approx 50~000 \approx 8000~\mathrm{yr}$).
Note that in order to better compare the two sample times, the $x$-axis in Fig.~\ref{fig.to} does not 
display the start time $t_0$ of the sampling, but rather the centre, $t_0+N\tau/2$ (i.e. each point in t
he graph lies at the temporal centre of the sample).

For region C and long sampling times, the dimension is found to
approximately  be  $\nu_{\text{C}}\approx5$. For shorter sampling times,
it is generally close to five, though occasionally dipping down
to  $\nu=3$. This indicates that on the considered length scale
the  dynamics  are  in  general  five-dimensional, but can also
become  three-dimensional  over  short intervals of $\approx 1000~\text{yr}$.
For  region  B  and  $t_0\lesssim60~000$, the dimension strongly  fluctuates  between  6  and  10,  whilst  for $60~000 \lesssim t_0 \lesssim 150~000$ it becomes stable around 7.
This is seen for both short and long sampling times.

\subsection{Derivative systems}

To aid in interpreting  the  above results we examine derivative systems of  Trappist-1.

\begin{figure}
\includegraphics[width=\linewidth]{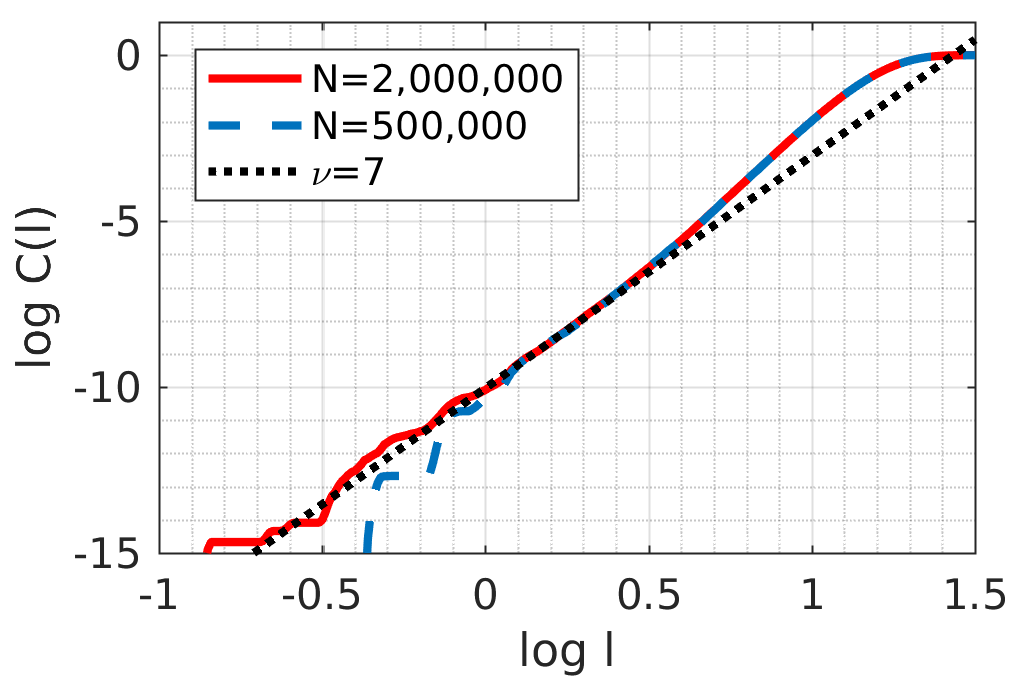}
\caption{
\label{fig.m0}
Double logarithmic plot of $C(l)$ over $l$ for the Trappist-1 system with non-interacting planets, with $t_0=0$ and $\tau=0.01$.
}
\end{figure}

Consider first, non-interacting planets.
The initial position of the trajectory at $t=0$ is the same as before [i.e. the coordinates and momenta taken from~\citet{TamayoReinPetrovichMurray2017}], but the subsequent  dynamics are calculated by setting the 
interactions between the planets to zero and only keeping the interaction with the central star.
Since the system is now one of seven independent planets -- each forming an independent, one-dimensional Kepler problem -- one can expect a total dimension of seven.
The dimension is calculated for two different sizes $N$ of the set $\{\mathbf{X}_n\}$, $N=500~000$ and $N=2~000~000$ and
the time-delay between subsequent points is set to $\tau=0.01$.
The result is shown in Fig.~\ref{fig.m0}.
One can fit a curve with the expected slope of $\nu=7$ to the data for $0.0<\log l<0.5$.
However, the data is seen to be rather noisy for $\log l < 0.2$ (even for $N=2~000~000$), and thus values from $\nu=\text{6 -- 8}$ would be justifiable as well.
This suggests that, in the previous calculations as well, an error of $\pm1$ 
should be assumed for the dimension of the Trappist-1 system calculated by this method.

As a second test, ensembles of slightly perturbed trajectories are considered.
Here, the perturbation consists of a slight change of the coordinate and momenta at the beginning
$t_0$ of the sampling time.
In particular, each coordinate/momentum of the reference trajectory is multiplied by a random number
 $r\in[1-\Delta,1+\Delta]$ at time $t_0$.
Any uniform movement of the centre of mass introduced by the random perturbation of the momenta was subsequently removed prior to applying the correlation dimension algorithm.

\begin{figure}
\includegraphics[width=\linewidth]{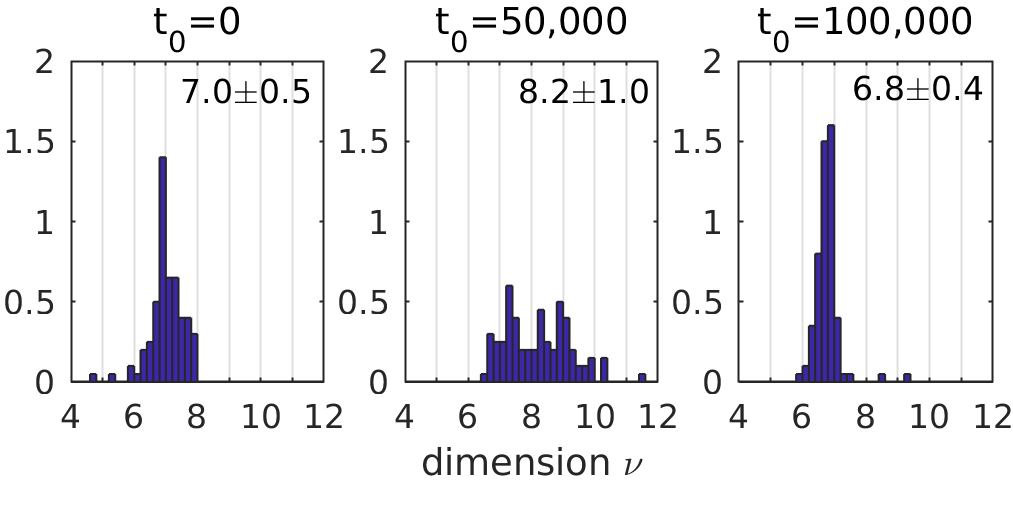}
\caption{
\label{fig.B}
Distribution of the dimensions in region B for an ensemble of perturbed trajectories with trajectories perturbed at the start $t_0$ of the sampling.
The perturbation is random and on the order of $\Delta=10^{-5}$ compared to the reference trajectory.
Also given are the mean value and standard deviation for each distribution.
}
\end{figure}

\begin{figure}
\includegraphics[width=\linewidth]{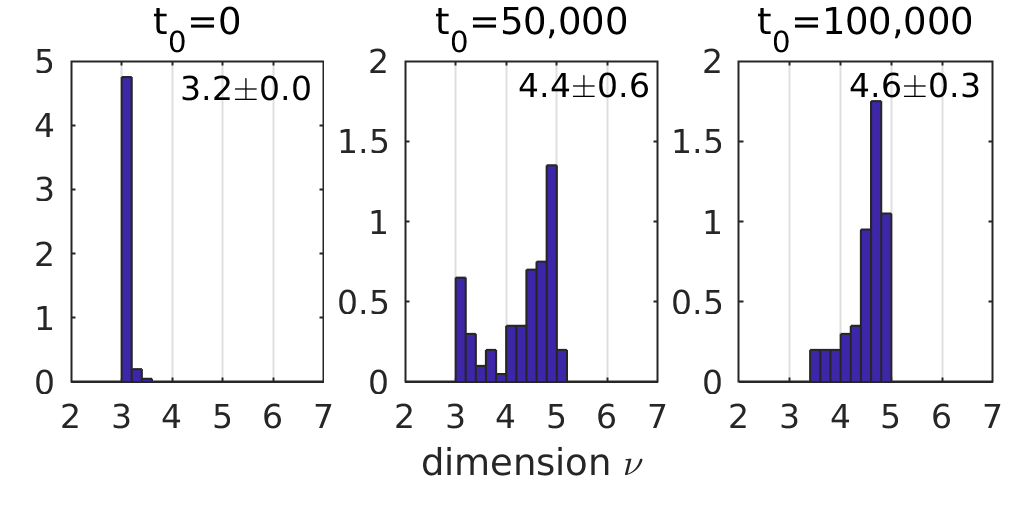}
\caption{
\label{fig.C}
Distribution of the dimensions in region C for an ensemble of trajectories with trajectories perturbed at the start $t_0$ of the sampling.
The perturbation is random and on the order of $\Delta=10^{-5}$ compared to the reference trajectory.
Also given are the mean value and standard deviation for each distribution.
}
\end{figure}

In Fig.~\ref{fig.B} the result of a weak perturbation of $\Delta=10^{-5}$ on region B is shown.
In particular, for three initial times $t_0$, an ensemble of 100 perturbed trajectories was created
and the dimension calculated for each perturbed trajectory, using $N=500~000$ and $\tau=0.01$ 
as numerical parameters.
For $t_0=0$ and $t_0=100~000$, the distribution of the dimensions is seen to peak at at $\nu\approx7$.
For $t_0=50~000$, a higher dimension with a larger width of the distribution is found: $\nu=8.2\pm1.0$.

In Fig.~\ref{fig.C}, the effect of the weak perturbation on region C is shown, using the same numerical parameters.
All perturbed trajectories are seen to have a dimension between 3 and 5 with most trajectories closer to $\nu=5$.
Thus the perturbation does not change the nature of the structures in region  since the
dimension of the reference trajectory in region C lies between 3 and 5.
Remarkably, for $t_0=0$, the perturbed trajectories show the same low dimension of $\nu\approx3$ as does the reference trajectory.

\begin{figure}
\includegraphics[width=\linewidth]{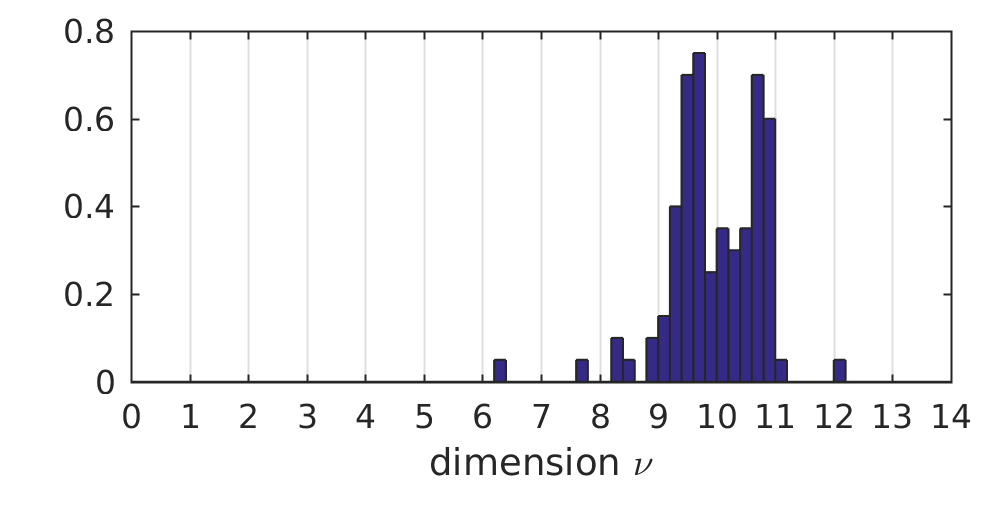}
\caption{
\label{fig.largeDelta}
Distribution of the dimensions for an ensemble of trajectories with larger 
perturbed initial conditions, $\Delta=10^{-3}$.
The deviations from the Trappist-1 system are so large
that for most cases the different regions are lost and only one region with a linear slope remains;
this slope is presented here.
The mean and standard deviation of the distribution are $10.0\pm0.8$.
}
\end{figure}

Finally, a larger perturbation of $\Delta=10^{-3}$  was considered.
In this case, and for most trajectories, the correlation sum $C(l)$ no longer divides into two regions B
 and C of different linear slopes; only one linear region is found.
The distribution of the dimensions is shown in Fig.~\ref{fig.largeDelta} for $t_0=0$:
For most perturbed trajectories, the dimension is higher than for the reference trajectory, 
and there is a broad distribution of dimensions: $\nu=10.0\pm0.8$.

\subsection{Discussion}

Region B lies in the interval $0.3 \lesssim l \lesssim 1$, and thus describes phase-space structures smaller than a planetary orbit (for the applied scaling a single planet's orbit has the extension of approximately 1 in phase-space).
Most of the times sampled, gave a dimension of space in which the system moves of $\nu_{\text{B}}=7$.
It is noteworthy that Trappist-1 therefore has the same dimension as a system of non-interacting planets on a size-scale in phase-space smaller than a planets' orbit, despite the fact that the planets are strongly interacting.

A remarkable feature of region B is the fact that the dimension $\nu_{\text{B}}$ increases from $\nu_{\text{B}}=7$ at $t_0=0$ to $\nu_{\text{B}}=10$ around $t_0=40~000$, but eventually returns to $\nu_{\text{B}}=7$ for $t_0>60~000$.
Since weak perturbations of the trajectory at $t_0=0$ and $t_0=100~000$ do not lead to an increase of the dimension (see Fig.~\ref{fig.B}), it is likely that this decrease of the number of integrals of motion around $t_0\approx40~000$ is real and not due to a numerical error.
A possible explanation for this behaviour is that the system occupies highly stable islands in the phase-space at $t_0=0$ and $t_0>60~000$, and migrates between these islands.
To test this explanation, a Fourier analysis of the eccentricities of the seven planets was done.
This analysis revealed highly coupled dynamics, which obscured possible evidence for the migration theory.
A more detailed analysis, beyond the scope of this work, will be necessary to reveal the origin of the change of the number of integrals of motion.

Region C lies within the interval $1\lesssim l \lesssim 2$ so that the phase-space structures it describes are larger than a single planet's orbit.
If the planets would move uncorrelated, the smallest dimension to be found should be seven; the value of five in region C indicates that the dynamics of two pairs or a triplet of planets are correlated.
This correlation could in principle also occur for non-interacting planets if the orbital periods coincidentally fit; however, to yield the observed results (5-dimensional structures stable over at least 100~000 orbital periods), the coincidental resonance would have to be exact to at least 10~ppm, and that for at least two pairs of planets.
It seems thus likely that the low dimension in region B is due to the interaction of the planets.

Considering the more strongly disturbed trajectories (see Fig.~\ref{fig.largeDelta}), the results indicate that the stability and high number of integrals of motion found for the trajectory of Trappist-1 may not be typical, but a specific feature of the Trappist-1 planetary system; i.e., the system may occupy a part of phase-space that is stable regardless of, or even due to, the interactions between planets.
This is consistent with the results of~\citet{TamayoReinPetrovichMurray2017}.
If the observed orbital parameters constrain the dimension of the system well enough, then this could be further evidence supporting the idea that the Trappist-1 system formed via convergent migration.
Note however that in the present study only a small region of phase-space was investigated, and thus the results are only an indication of this possibility.

\section{Conclusion}

This article presents a study of the phase-space dimension of the dynamics of the Trappist-1
planetary system.
This system consists of seven planets orbiting a dwarf star with
the orbits sufficiently close that the planets are strongly interacting.
In particular, an adapted version of the correlation dimension 
method~\citep{grassberger83a,grassberger83b} was used to determine the dimension of the surface that a trajectory of the system traces out in phase-space.
This dimension then provides the number of integrals of motion in the region traced out by the trajectory.

The  central result  is that the Trappist-1 planetary system does not show 38-dimensional dynamics, as is possible for an eight-body problem, on the investigated  length-~and time-scales.
Instead, on the size scale of a single  orbit,  the  system was found to have the same number of integrals of motion  as  would  be  expected  if  the seven planets were non-interacting.
Further tests showed that the trajectory is possibly located in a small and very  stable  region  of phase-space.
Fluctuations of the number of integrals due to the system's  dynamics were not able to bring the system out of the
stable  region;  random  perturbations on the order of 0.1\% on
the    other   hand   were.   Furthermore,   there   are   even
lower-dimensional  structures,  indicating  that  some  of  the
orbits are partially locked.

It should be noted that the time-scale considered in this study 
is $\sim$20~000 years, shorter than the shortest time of
0.5~Myr  for  which instabilities of the Trappist-1 system were recently
reported~\citep{gillon17}.  A  future  study on the dimension for
these  time  scales  is  therefore  of interest; in particular, 
chaotic dynamics may start to dominate and thus increase the number
of dimensions.

\section*{Acknowledgements}
We thank Professor Itamar Procaccia and Mr. Daniel Tamayo for helpful discussions.
This research has been supported by the NSERC Discovery Grants to P.B. and H.R.
Part of the computations were performed on the GPC supercomputer at the 
SciNet HPC Consortium. SciNet is funded by: the Canada Foundation for Innovation under the auspices of Compute Canada; the Government of Ontario; Ontario Research Fund - Research Excellence; and the University of Toronto.

\bibliographystyle{mnras}
\bibliography{bibliography_trappist}

\bsp	
\label{lastpage}
\end{document}